\documentclass[a4paper,11pt]{article}
\pdfoutput=1 

\usepackage{jcappub} 

\usepackage[T1]{fontenc} 
\usepackage{subfig}
\usepackage{float}

\usepackage{changes}

\title{\boldmath An updated view and perspectives on high-energy gamma-ray emission
from SGR J1935+2154 and its environment}

\author[a,b,c,1]{Jaziel G. Coelho\note{Corresponding author.},}
\author[d]{Luana N. Padilha,}
\author[d,e,f]{Rita C. dos Anjos,}
\author[d]{Cynthia V. Ventura,} 
\author[b,g]{and Geanderson A. Carvalho}

\affiliation[a]{N\'ucleo de Astrof\'isica e Cosmologia (Cosmo-Ufes) \& Departamento de F\'isica, Universidade Federal do Esp\'irito Santo, 29075--910, Vit\'oria, ES, Brazil}
\affiliation[b]{Departamento de F\'isica, Universidade Tecnol\'ogica Federal do Paran\'a, 85884-000, Medianeira, PR, Brazil}
\affiliation[c]{Divis\~ao de Astrof\'isica, Instituto Nacional de Pesquisas Espaciais, Avenida dos Astronautas 1758, 12227-010, S\~ao Jos\'e dos Campos, SP, Brazil}
\affiliation[d]{Departamento de F\'isica, Universidade Tecnol\'ogica Federal do Paran\'a, 82590-300, Jardim das Americas, Curitiba, PR, Brazil}
\affiliation[e]{Departamento de Engenharias e Exatas, Universidade Federal do Paran\'a (UFPR), Pioneiro, 2153, 85950-000 Palotina, PR, Brazil}
\affiliation[f]{Applied Physics Graduation Program, Federal University of Latin-American Integration, 85867-670, Foz do Igua\c{c}u , PR, Brazil}
\affiliation[g]{Instituto de Pesquisa e Desenvolvimento (IP\&D), Universidade do Vale do Para\'iba, 12244-000, S\~ao Jos\'e dos Campos, SP, Brazil}

\emailAdd{jaziel.coelho@ufes.br}
\emailAdd{luanapadilha@alunos.utfpr.edu.br}
\emailAdd{ritacassia@ufpr.br}
\emailAdd{cynthiaventura@utfpr.edu.br}
\emailAdd{gacarvalho@utfpr.com}

\abstract{
SGR J1935+2154 was discovered in 2016 and is currently one of the most burst-active Soft Gamma-ray Repeaters (SGR), having emitted many X-ray bursts in recent years. In one of our previous articles, we investigated the contribution to high-energy and very high-energy gamma-ray emission (VHE, $E > 100$~GeV) due to cosmic-ray acceleration of SNR G57.2+0.8 hosting SGR J1935+2154 using the GALPROP propagation code. However, follow-up observations of SGR 1935+2154 were made for 2 hours on April 28, 2020, using the High Energy Stereoscopic System (H.E.S.S.). The observations coincide with X-ray bursts detected by INTEGRAL and Fermi/Gamma-ray Burst Monitor (GBM). These are the first high-energy gamma-ray observations of an SGR in a flaring state, and upper limits on sustained and transient emission have been derived. Now that new H.E.S.S. observations have been made, it is interesting to update our model with respect to these new upper limits. We extend our previous results to a more general situation using the new version of GALPROP. We obtain a hadronic model that confirms the results discussed by H.E.S.S.. This leads to an optimistic prospect that cosmic ray gamma rays from SGR J1935+2154 can contribute to the overall gamma energy density distribution and in particular to the diffusion gamma rays from the Galactic center.}

\begin{document}
\maketitle
\flushbottom

\section{Introduction}

The Soft Gamma-ray Repeater (SGR) J1935+2154 was initially detected by the BAT (Burst Alert Telescope) instrument aboard the Swift satellite as an X-ray burst. Subsequent observations of this source allowed to classify it as a magnetar and they found that the source became active again in April 2020, whilst it exhibited multiple and severe X-ray burst activity (see ~\cite {2016MNRAS.457.3448I,2020ApJ...898L..29M,2020ApJ...902L...2B} and references therein). SGRs are a diverse set of sources with huge magnetic fields of the order of $10^{14}-10^{15}$ G, rotational periods $P\sim(2-12)$ s, slowing down rates $\dot{P}\sim(10^{-15}-10^{-10})$  s/s, persistent X-ray luminosity as large as $10^{35}$ erg/s, transient activity in the form of outbursts of energy around $10^{41}-10^{43}$ erg, and, for some sources, the presence of giant flares, whose
typical luminosities are $10^{44}-10^{47}$ erg (see, e.g., ~\cite{2014ApJS..212....6O} and references therein). 
The emission nature of SGRs remains a cause for debate, and several scenarios were proposed to explain their observed spectra and properties~\citep[see][for detailed reviews]{2008A&ARv..15..225M, 2015RPPh...78k6901T, 2017ARA&A..55..261K}. 
Examples thereof are magnetars~\cite{1992ApJ...392L...9D,Thompson1995,2007ApJ...657..967B,2012ApJ...748L..12R,2013ApJ...762...13B,2020ApJ...889..165D}, accreting neutron stars~\cite{1995A&A...299L..41V,2001ApJ...554.1245A}, rotation-powered pulsars~\cite{2017A&A...599A..87C}, quark stars \cite{2011MNRAS.415.1590O}, or massive, fast-spinning and highly magnetized white dwarfs ~\cite{Malheiro2012,Coelho2014c,doi:10.1142/S021827181641025X,2017MNRAS.465.4434C,2020ApJ...895...26B,2020MNRAS.498.4426S}.  

On April 28, 2020, a very bright radio outburst of SGR J1935+2154 was identified that turned out to be brighter than any radio burst seen from any galactic source to date. An extremely bright millisecond-duration radio burst was emitted by SGR J1935+2154 and detected with the Canadian Hydrogen Intensity Mapping Experiment (CHIME) and STARE2 telescopes~\cite{2020Natur.587...54C,2020ATel13684....1B}. The radio burst turned out to be temporally
coincident with a bright hard X-ray burst detected with
the NICER \cite{2020ApJ...904L..21Y}, INTEGRAL \cite{2020ApJ...898L..29M}, Konus-Wind~\cite{2021NatAs...5..372R}, Insight Hard X-ray Modulation Telescope - HXMT~\cite{2021NatAs...5..378L} and AGILE satellites \cite{2021NatAs...5..401T}.


Fast radio bursts are extremely intense millisecond-long radio pulses of primarily extragalactic origin and showcasing the characteristic dispersion sweep of radio pulsars. The physical nature of these bursts is unknown, and a variety of explanations have been considered, including synchrotron maser emission from young SGRs in supernova remnants (SNRs)~\citep[see][and references therein]{2020Natur.587...45Z} and magnetospheric emission~\citep[see][for detais]{2020MNRAS.498.1397L}. SGRs have been proposed as sources of FRBs~\citep[see][]{2010vaoa.conf..129P,2014MNRAS.442L...9L,2019MNRAS.485.4091M,2020ApJ...896..142B}. In particular, \cite{2017ApJ...843L..26B} suggest that repeating FRBs
are generated not far from the surface of the source, as a result of
ultrarelativistic internal shocks in the magnetar wind, which are launched by the magnetospheric flares. 

In one of our previous articles we have obtained the contribution to the high energy and very high-energy gamma-ray (VHE, $E > 100$ GeV) emission due to cosmic-ray acceleration from SNR G57.2+0.8 hosting SGR J1935+2154 with the use of the GALPROP propagation code\footnote{\url{ http://galprop.stanford.edu}}~\citep[see][]{2021JCAP...10..023D}. We consider the SGR J1935+2154 and its
environment as a single source near to the Galactic center. Then, we have proposed that the above setting can provide a more comprehensive scenario for the generation of GeV-TeV gamma-rays. In fact, the diffuse gamma-ray emission from the Galactic plane was first detected by EGRET \cite{Hunter_1997} and then followed by Fermi LAT measurements~\cite{Fermi2012}. Additionally, a TeV diffuse emission originated in the central part of our Galaxy was detected by ground-based imaging atmospheric Cherenkov telescopes~\citep{HESS2006, Archer_2016, Magic2020}. 
The observations by the High Energy Stereoscopic System (H.E.S.S.) of a gradient in the cosmic-ray (CR) profile derived from the diffuse VHE gamma-ray emission, with a peak at the inner regions, imply an injection by a steady source located at the center of the Milk Way~\cite{HESS2016}.

The main goal of the present paper is to extend our previous study where we have investigated particle acceleration models for SNR G57.2+0.8 hosting SGR J1935+2154~\citep[see][for details]{2021JCAP...10..023D} motivated by H.E.S.S., Fermi-GBM and INTEGRAL VHE gamma-ray observations~\citep[see][]{2020ApJ...902L..43L,2021ApJ...919..106A}. Here, we have used GALPROP propagation code~\cite{Strong_1998, Porter_2017, J_hannesson_2018} in its enhanced version~\cite{2021arXiv211212745P} performing 3D diffusion/re-acceleration model simulations. Further, we use the new flux from the pulsar spin-down timescale for SGR J1935+2154. It is assumed that the SGR J1935+2154 is injecting accelerated particles into the interstellar medium (ISM) in equal numbers with a
fraction of its spin-down power converted to pairs. After
injection, the particles propagate via a diffusive process~\citep[see e.g.,][]{2019ApJ...879...91J}. The mechanism we demonstrate here is an indication that improves the understanding of the leptonic and hadronic origins of the gamma-ray emission. Also, such analysis would show whether the gamma-ray emission around this region is due to leptonic or hadronic processes. H.E.S.S. measurements at VHE - TeV energies of SGR J1935+2154 are crucial for constraining SGR particle acceleration scenarios up to the knee. As a result, they may provide independent properties of the magnetospheric regions of SGRs and hence unveil some of the physics taking place in their outermost regions.
To do so this paper is organized as follows. Section 2 is
devoted to an update and brief summary of H.E.S.S. observations. In Section 3 and Section 4 we present how the
calculations are done and discuss the obtained results.

\section{High-energy gamma-ray emission
from SGR
J1935+2154}\label{environment}

Follow-up observations of SGR 1935+2154 were carried by H.E.S.S. for 2 hr on 2020 April 28. The observations are coincident with X-ray bursts detected by INTEGRAL and Fermi-GBM, thus providing the first very high energy gamma-ray observations of a SGR in a flaring state. High-quality data acquired during these follow-up observations allow us to perform a search for short-time transients. No significant signal at energies $E> 0.6$ TeV is found, and upper limits on the persistent and transient emission were derived~\citep[see][for details]{2021ApJ...919..106A}.

The integral upper limits derived from the 2 hr of H.E.S.S. observation, assuming a spectral index of $E^{-2.5}$, can be translated into upper limits on the flux $F(E > 600$ GeV) $<2.4\times10^{-12}$ $\rm erg$ $\rm cm^{-2}s^{-1}$. SGR 1935+2154 is associated with the middle-aged galactic SNR G57.2+0.8 at a distance of about 6.6 kpc  (see \cite{2021JCAP...10..023D,2020ApJ...905...99Z} and references therein). Assuming this distance, \cite{2021ApJ...919..106A} derive a luminosity upper limit $L(E > 600$ GeV) $<1.3\times10^{34}$ erg/s. This places constraints on persistent VHE emission from SGR 1935+2154 during the H.E.S.S. observations. 

Moreover, the simultaneous X-ray emission and radio bursts offered the missing link to correlate FRBs with SGRs. The X-ray emission can be an indication that protons and electrons are being accelerated at TeV scales via interactions with matter in the region or through inverse Compton scattering, respectively, and H.E.S.S. observations can help better understand this scenario. As well discussed in \cite{2021JCAP...10..023D}, if SGRs are cosmic-ray sources at the TeV-PeV energy range, they could contribute to the diffuse gamma-ray emission in the Galaxy given that it is expected to originate from the interactions of cosmic rays with Galactic interstellar gas and radiation fields. It is also worth noting that H.E.S.S. has shown a strong correlation between the Central Molecular Zone (CMZ) and the brightness distribution of diffuse emission, possibly indicating that these gamma rays originate from VHE protons interacting with matter in these regions. It has been argued that the origin of these CRs via Inverse Compton scattering is likely prevented by the strong radiative losses of TeV electrons that limit their propagation in the CMZ~\cite{HESS2018}. Moreover, H.E.S.S. has observed a high-energy CR density in the CMZ that is an order of magnitude larger than in the rest of the Galaxy and that could be interpreted by the existence of more than one accelerating source in the Galactic center~\cite{HESS2018}. This clearly motivates studies considering G57.2+0.8 and/or SGR J1935+2154 as such sources.
Recently, the Tibet AS$\gamma$ collaboration reported the first detection of sub-PeV diffuse gamma rays in the Galactic disk. The lack of correlation between the observed events and known TeV sources virtually rules out a leptonic origin for this diffuse emission, and argues for an origin by hadrons accelerated in this region at PeV energies \citep[see][for details]{Tibet2021}.

\section{Simulations}\label{model}

In this section, we briefly review the main features of the model to highlight the relevant improvements of our simulations compared with previous results reported in~\cite{2021JCAP...10..023D}. 
In order to assess the distribution of gamma rays from the propagation of cosmic rays (CRs) for SNR G57.2+0.8, which hosts SGR J1935+2154, we used GALPROP code~\cite{Strong_1998, Porter_2017, J_hannesson_2018} in its latest version v57~\cite{2021arXiv211212745P} performing 3D diffusion/re-acceleration model simulations. The corresponding spectra of gamma rays reaching the Earth taking into account the propagation effects and interactions with background radiation were determined. For a given source distribution and boundary conditions, GALPROP solves the transport equation. Energy loss, fragmentation and decay, convection, diffusion, and re-acceleration processes are included in the simulations. Based on the input CR source abundances, GALPROP computes a comprehensive network of isotope production. A second-order implicit Crank-Nicholson approach is used for the numerical solution. The spatial boundary conditions are based on the assumption that free particles escape. The diffusion coefficient for a given halo size is defined by the B/C nuclear ratio as a function of rigidity and dispersion parameter~\cite{Strong_1998, Porter_2017, J_hannesson_2018}. To a better comprehension, in this paper we improve the model described in~\cite{2021JCAP...10..023D} with its environment and GALPROP parameters, used to better describe the SNR G57.2+0.8 hosting SGR J1935+215.

However, the calculation of the spectral energy distribution of the total gamma-ray emission uses parameters of the source SGR J1935+2154. The modelling assumes that accelerated particles are injected into the ISM and the spectral model for the injected particles is obtained with a power law $dq(p)/dp \propto p^{-\alpha}$, where $\alpha$ is the spectral index and $q(p)$ the injection energy spectrum and, the total spectrum is normalized so that the total injected power is given by the expression~\cite{2009PhRvD..80f3005M,2019ApJ...879...91J,2021arXiv211212745P}
\begin{equation}
L = L(t) + L' \ \ \  \mathrm{and} \ \ \  L(t) = \eta L_{0}\Biggl(1 + \frac{t}{\tau_0}\Biggl)^{-2} ,
\label{power}
\end{equation}

where ${L_0}$ is the initial spin-down power of the SGR, 
$\eta = 1.0$ is the efficiency factor,
$\tau_0 = 3.37$ $\mathrm{yr}$ is the pulsar time scale defined as the ratio of the initial rotational energy to the initial spin-down luminosity~\citep[see][]{2009PhRvD..80f3005M} and $L'$ is the gamma-rays luminosity of the burst state from H.E.S.S. upper limits \cite{2021ApJ...919..106A}. The initial spin-down power is calculated using the current spin-down power of $1.7\times 10^{34}$ $\mathrm{erg\;s^{-1}}$ assuming the characteristic age of $3.6\times 10^{3}$ $\mathrm{yr}$\footnote{In the magnetic dipole approximation, we can use $\Omega/\dot{2\Omega}=-(t+\tau)$ \citep[see Eq. A5 of][]{2009PhRvD..80f3005M}, the characteristic age $t$, and the current values of $\Omega$ and $\dot{\Omega}$ to calculate that its pulsar time scale $\tau_0 = 3.37$ yr. Assuming NSs canonical parameters (moment of inertia $I\sim1.4\times10^{45}$ $\rm g cm^{2}$) we can derive the initial rotational energy. The similar approach was developed by \citep{2009PhRvD..80f3005M} to estimate the initial energy by using the current spin-down luminosity~\citep[see also][]{2019ApJ...879...91J,2021arXiv211212745P}.}. The luminosity $L$ represents the total emission of particles from SGR J1935+2154. We assume that this emission is caused by the rotational energy loss of the pulsar ~\citep[see e.g.,][]{2009PhRvD..80f3005M}, and by the luminosity in a flaring state from the magnetar. Then, we have considered
the magnetic energy injection rate (for both quiescent and flaring/outburst emissions). The distance of the SGR J1935+2154 is $6.6\pm0.7$ kpc~\citep{2011A&A...536A..83S,2013ApJS..204....4P,2020ApJ...905...99Z}.

Calculations are performed in a Cartesian grid, with the Galactic plane assumed to be the X-Y plane at whose origin the GC is located. The Galactic volume spans up to 18 kpc in the X and Y directions, with a halo height (h) of 8 kpc in the X-Y direction. We use the treatment of inhomogeneous diffusion properties in the space around CR sources, modelling two-zone diffusion scenarios, as described in~\cite{2021JCAP...10..023D}.

The generation of $\gamma$-rays is determined by the propagating CR distributions, including primary, secondary, and inelastically scattered protons. Inverse Compton scattering is calculated by the interactions between anisotropic background photon distribution with a Galactic interstellar radiation field. As well as different contributions of the pion-decay and bremsstrahlung are calculated from emission levels using the column densities of $H$ gas. Models of the Galactic magnetic field are used to determine the synchrotron emission. The $\gamma$-ray and synchrotron sky maps result from integrating the relevant emissivities with gas distributions, interstellar radiation, and magnetic fields~\citep[see][]{Strong_1998}. The gamma-ray flux is calculated from G57.2+0.8 hosting SGR J1935+2154 and compared with those published in \cite{2021JCAP...10..023D}.

\section{Discussions}\label{model}

Gamma-ray emission can be produced by hadronic and leptonic processes triggered by the interaction of CRs accelerated in the molecular cloud region and ISM. We simulate the CR flux as a result of SGR J1935+2154 injecting particles into the surrounding space and account the contribution of gamma-ray emission by the propagation of CRs for SNR G57.2+0.8. 

\begin{figure*}[h]
  \centering
   \subfloat[$\alpha = 2.2$]{\includegraphics[angle=0,width=0.5\textwidth]{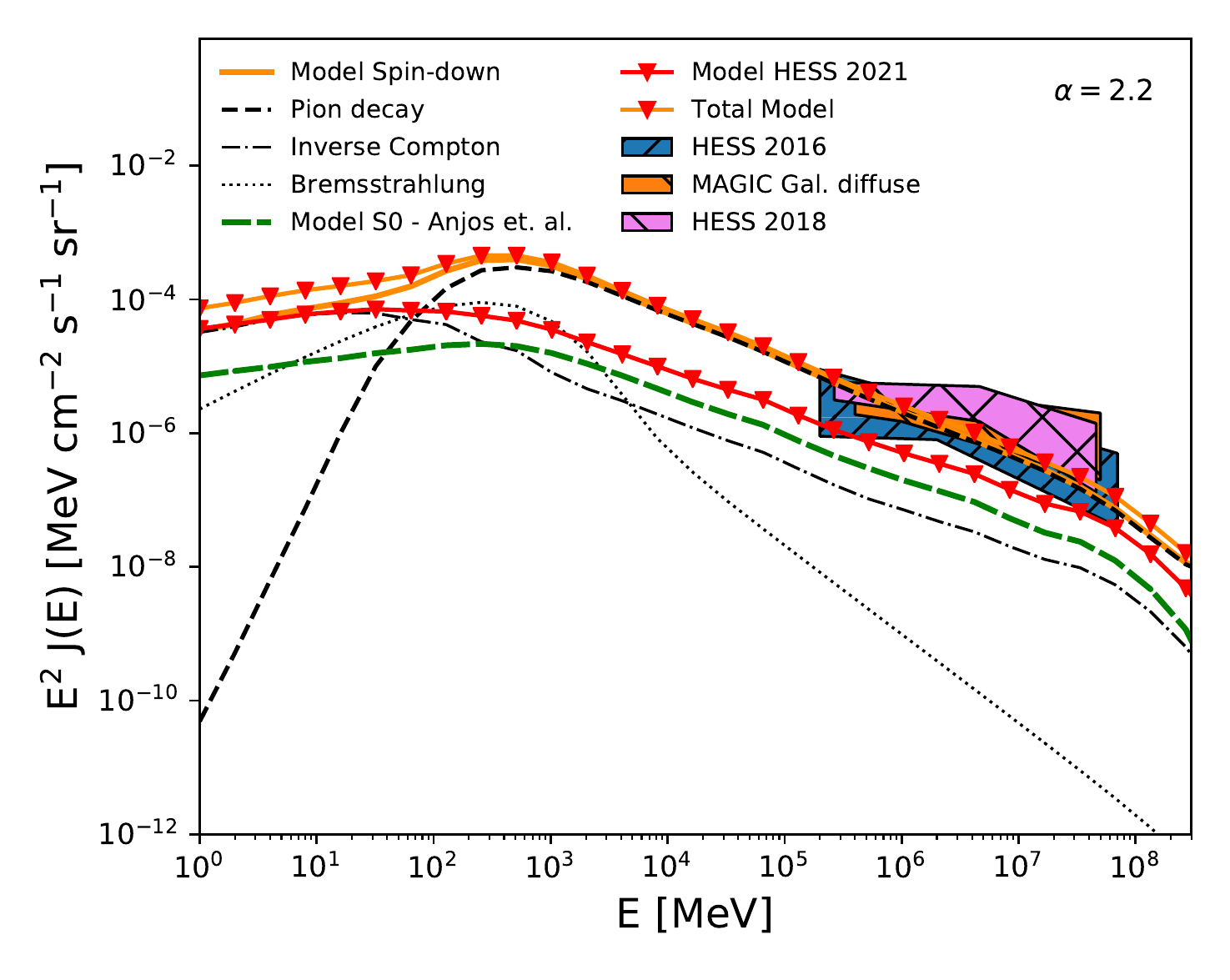}}
    \subfloat[$\alpha = 2.4$]{\includegraphics[angle=0,width=0.5\textwidth]{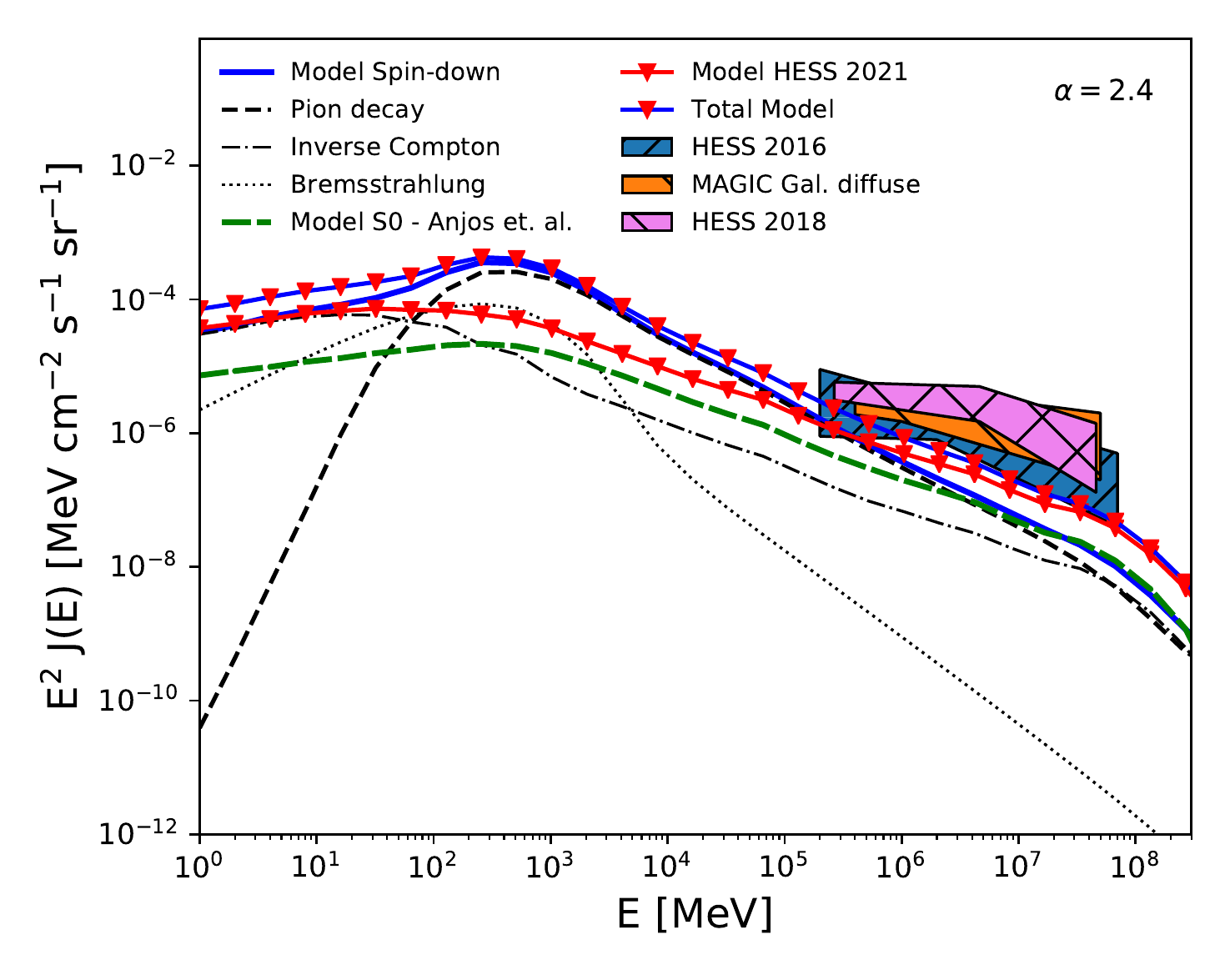}}\\
   \subfloat[$\alpha = 2.6$]{\includegraphics[angle=0,width=0.5\textwidth]{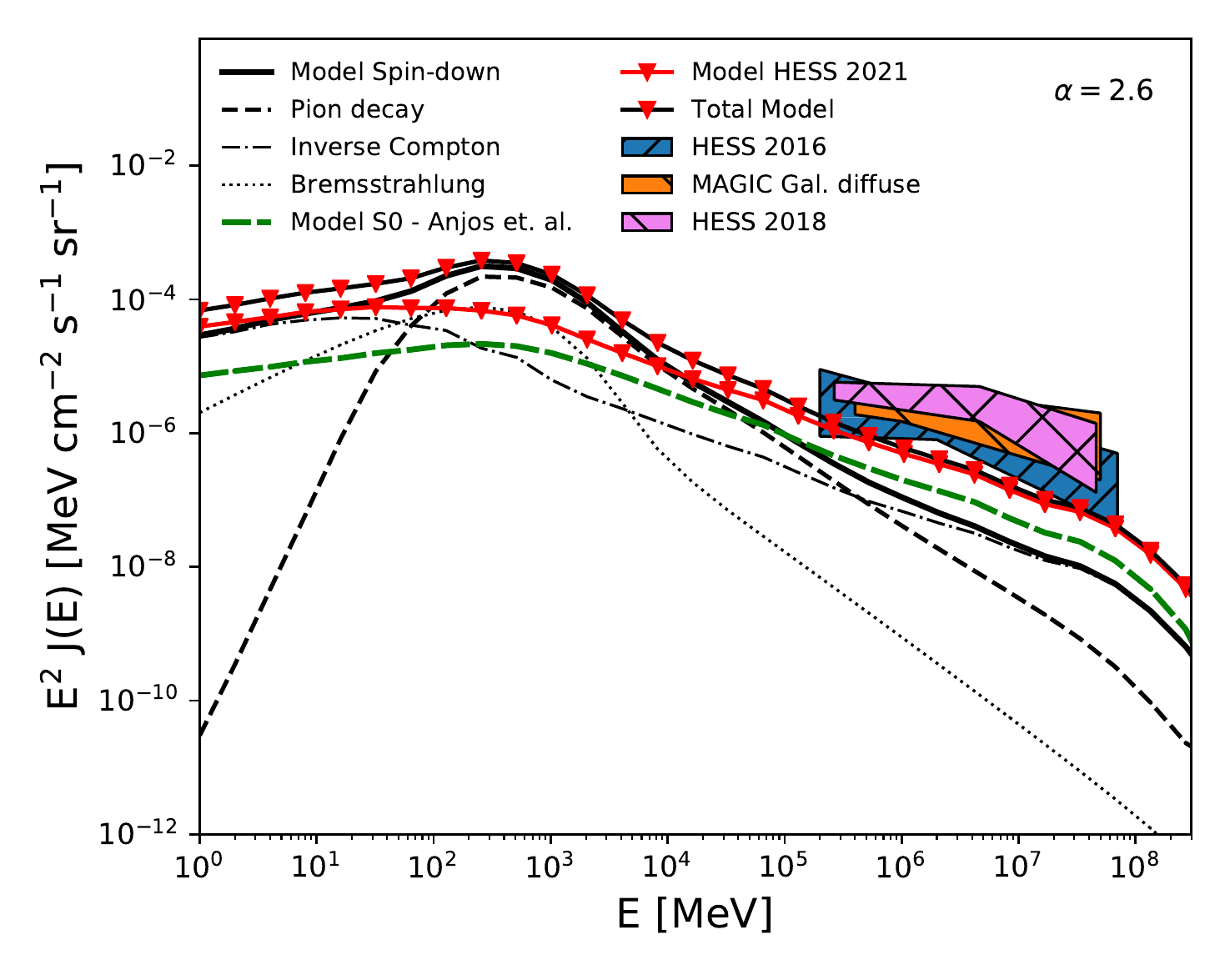}}
   \caption{Spectral energy distribution of the gamma-ray emission. The total gamma-ray spectrum is the sum of pion decay, inverse Compton, and bremsstrahlung from the spin-down model, $L(t)$, and the H.E.S.S. 2021 model, described from the flaring state $L'$, of the SGR J1935+2154. See text for details. The models use the 2D gas distribution~\cite{J_hannesson_2018}.}
    \label{gamma_models}
\end{figure*}

In order to ﬁx the same environment as in~\cite{2021JCAP...10..023D} we choose diffusion constant $D_{0}= 2.5\times 10^{28}$ $\mathrm{cm^{2}\;s^{-1}}$, slope of the diffusion coefficient $\delta = 0.6$, Alfvén velocity $v_{a} = 28.0$ $\mathrm{km\;s^{-1}}$ and spectral slope $\alpha = 2.2, 2.4, 2.6$. The results are described in Figure~\ref{gamma_models}. The Figure displays the behavior of the $S0$ model described in~\cite{2021JCAP...10..023D}, in which the VHE gamma-ray emission from SNR G57.2+0.8 and SGR J1935+2154 was normalized as the sum of pion decay and inverse Compton spectra, with an upper limit on the integral flux of TeV gamma rays from H.E.S.S. Observatory up to 99.5\% CL~\cite{HESS2018}. The model presented in \cite{2021JCAP...10..023D}, although consistent, shows a lower flux than the current model. This is because the flux calculation in previous work is associated with the upper limits of the gamma integral measured by H.E.S.S.~\cite{HESS2018}, which was updated in \cite{2021ApJ...919..106A}. 

Figure~\ref{gamma_models} shows the data of diffuse TeV energy gamma-ray emission from the Galactic Center of H.E.S.S. \cite{HESS2016,HESS2018} and MAGIC models \cite{Magic2020}. According to our findings, gamma-rays of CRs from SGR J1935+2154 can make contribution to the total gamma energy density distribution and, in special, to the diffusion gamma ray from Galactic Center. Up to $10^{4}$ $\mathrm{MeV}$, the pion emission component dominates. The differences in the spectral indices show that smaller indices make a larger contribution to the Galactic diffuse gamma, see Fig.~\ref{gamma_models} - (a,b,c). The figure~\ref{gamma_models} also describes the production of the VHE gamma-ray emission from particles following the method discussed in \cite{2021JCAP...10..023D}. The method makes a conservative connection between the upper limit of the integral flux of TeV gamma-rays detected by H.E.S.S. \cite{2021ApJ...919..106A} and the cosmic-ray luminosity at the source. The red line on figure~\ref{gamma_models} shows this result. We suggest that the flare was the result of the interaction of particles accelerated, producing $p-p$ collisions and generating gamma-ray emission and, therefore, dominating the VHE spectrum, figure~\ref{gamma_models}-(b,c). This emission is below than the quiescent one, described by the orange line, figure~\ref{gamma_models}-(a), because the spectral index is harder, consequently generating more pions at higher energies.

This simple model presented is an interesting window that improves the understanding about the hadronic origins of gamma-ray emission at high energies, as discussed by H.E.S.S.~\cite{2021ApJ...919..106A}. The approach presented here becomes more predictive considering the energy losses during the CRs propagation, magnetic field, distance and the spin-down age of SGR J1935+2154.

\acknowledgments
The authors thank the referee for comments which helped to
improve the quality of the manuscript.
J.G.C. is grateful for the support of CNPq (311758/2021-5) and FAPESP Project No. 2021/01089-1. The research of R.C.A. is supported by Conselho Nacional de Desenvolvimento Cient\'{i}fico e
Tecnol\'{o}gico (CNPq), grant number 310448/2021-2. R.C.A. and L.N.P thank for the support of L'Oreal Brazil, with partnership of ABC and UNESCO in Brazil. We acknowledge the National Laboratory for Scientific Computing (LNCC/MCTI, Brazil) for providing HPC resources of the SDumont supercomputer, which have contributed to the research results reported within this paper. URL: http://sdumont.lncc.br. J.G.C and R.C.A acknowledge the financial support of FAPESP Project No. 2021/01089-1. The authors acknowledge the financial support of NAPI “Fenômenos Extremos do Universo” of Fundação Araucária.

\bibliographystyle{ieeetr}
\bibliography{references}

\begin{thebibliography}{10}

\bibitem{2016MNRAS.457.3448I}
G.~L. {Israel}, P.~{Esposito}, N.~{Rea}, F.~{Coti Zelati}, A.~{Tiengo},
  S.~{Campana}, S.~{Mereghetti}, G.~A. {Rodriguez Castillo}, D.~{G{\"o}tz},
  M.~{Burgay}, A.~{Possenti}, S.~{Zane}, R.~{Turolla}, R.~{Perna},
  G.~{Cannizzaro}, and J.~{Pons}, ``{The discovery, monitoring and environment
  of SGR J1935+2154},'' {\em \mnras}, vol.~457, pp.~3448--3456, Apr. 2016.

\bibitem{2020ApJ...898L..29M}
S.~{Mereghetti} and et~al., ``{INTEGRAL Discovery of a Burst with Associated
  Radio Emission from the Magnetar SGR 1935+2154},'' {\em \apjl}, vol.~898,
  p.~L29, Aug. 2020.

\bibitem{2020ApJ...902L...2B}
A.~{Borghese}, F.~{Coti Zelati}, N.~{Rea}, P.~{Esposito}, G.~L. {Israel},
  S.~{Mereghetti}, and A.~{Tiengo}, ``{The X-Ray Reactivation of the Radio
  Bursting Magnetar SGR J1935+2154},'' {\em \apjl}, vol.~902, p.~L2, Oct. 2020.

\bibitem{2014ApJS..212....6O}
S.~A. {Olausen} and V.~M. {Kaspi}, ``{The McGill Magnetar Catalog},'' {\em
  \apjs}, vol.~212, p.~6, May 2014.

\bibitem{2008A&ARv..15..225M}
S.~{Mereghetti}, ``{The strongest cosmic magnets: soft gamma-ray repeaters and
  anomalous X-ray pulsars},'' {\em \aapr}, vol.~15, pp.~225--287, July 2008.

\bibitem{2015RPPh...78k6901T}
R.~{Turolla}, S.~{Zane}, and A.~L. {Watts}, ``{Magnetars: the physics behind
  observations. A review},'' {\em Reports on Progress in Physics}, vol.~78,
  p.~116901, Nov. 2015.

\bibitem{2017ARA&A..55..261K}
V.~M. {Kaspi} and A.~M. {Beloborodov}, ``{Magnetars},'' {\em \araa}, vol.~55,
  pp.~261--301, Aug. 2017.

\bibitem{1992ApJ...392L...9D}
R.~C. {Duncan} and C.~{Thompson}, ``{Formation of very strongly magnetized
  neutron stars - Implications for gamma-ray bursts},'' {\em \apjl}, vol.~392,
  pp.~L9--L13, June 1992.

\bibitem{Thompson1995}
C.~Thompson and R.~C. Duncan, ``{The soft gamma repeaters as very strongly
  magnetized neutron stars - I. Radiative mechanism for outbursts},'' {\em
  \mnras}, vol.~275, pp.~255--300, 7 1995.

\bibitem{2007ApJ...657..967B}
A.~M. {Beloborodov} and C.~{Thompson}, ``{Corona of Magnetars},'' {\em \apj},
  vol.~657, pp.~967--993, Mar. 2007.

\bibitem{2012ApJ...748L..12R}
N.~{Rea}, J.~A. {Pons}, D.~F. {Torres}, and R.~{Turolla}, ``{The Fundamental
  Plane for Radio Magnetars},'' {\em \apjl}, vol.~748, p.~L12, Mar. 2012.

\bibitem{2013ApJ...762...13B}
A.~M. {Beloborodov}, ``{On the Mechanism of Hard X-Ray Emission from
  Magnetars},'' {\em \apj}, vol.~762, p.~13, Jan. 2013.

\bibitem{2020ApJ...889..165D}
R.~C.~R. {de Lima}, J.~G. {Coelho}, J.~P. {Pereira}, C.~V. {Rodrigues}, and
  J.~A. {Rueda}, ``{Evidence for a Multipolar Magnetic Field in SGR J1745-2900
  from X-Ray Light-curve Analysis},'' {\em \apj}, vol.~889, p.~165, Feb. 2020.

\bibitem{1995A&A...299L..41V}
J.~{van Paradijs}, R.~E. {Taam}, and E.~P.~J. {van den Heuvel}, ``{On the
  nature of the `anomalous' 6-s X-ray pulsars},'' {\em \aap}, vol.~299, p.~L41,
  July 1995.

\bibitem{2001ApJ...554.1245A}
M.~A. {Alpar}, ``{On Young Neutron Stars as Propellers and Accretors with
  Conventional Magnetic Fields},'' {\em \apj}, vol.~554, pp.~1245--1254, June
  2001.

\bibitem{2017A&A...599A..87C}
J.~G. {Coelho}, D.~L. {C{\'a}ceres}, R.~C.~R. {de Lima}, M.~{Malheiro}, J.~A.
  {Rueda}, and R.~{Ruffini}, ``{The rotation-powered nature of some soft
  gamma-ray repeaters and anomalous X-ray pulsars},'' {\em \aap}, vol.~599,
  p.~A87, Mar. 2017.

\bibitem{2011MNRAS.415.1590O}
R.~{Ouyed}, D.~{Leahy}, and B.~{Niebergal}, ``{SGR 0418+5729 as an evolved
  Quark-Nova compact remnant},'' {\em \mnras}, vol.~415, pp.~1590--1596, Aug.
  2011.

\bibitem{Malheiro2012}
M.~Malheiro, J.~A. Rueda, and R.~Ruffini, ``{SGRs and AXPs as Rotation-Powered
  Massive White Dwarfs},'' {\em Publications of the Astronomical Society of
  Japan}, vol.~64, p.~56, 6 2012.

\bibitem{Coelho2014c}
J.~G. Coelho and M.~Malheiro, ``{Magnetic Dipole Moment of SGRs and AXPs
  Described as Massive and Magnetic White Dwarfs},'' {\em Publications of the
  Astronomical Society of Japan}, vol.~66, pp.~14--14, 2 2014.

\bibitem{doi:10.1142/S021827181641025X}
R.~V. Lobato, M.~Malheiro, and J.~G. Coelho, ``{Magnetars and white dwarf
  pulsars},'' {\em International Journal of Modern Physics D}, vol.~25, no.~09,
  p.~1641025, 2016.

\bibitem{2017MNRAS.465.4434C}
D.~L. {C{\'a}ceres}, S.~M. {de Carvalho}, J.~G. {Coelho}, R.~C.~R. {de Lima},
  and J.~A. {Rueda}, ``{Thermal X-ray emission from massive, fast rotating,
  highly magnetized white dwarfs},'' {\em \mnras}, vol.~465, pp.~4434--4440,
  Mar. 2017.

\bibitem{2020ApJ...895...26B}
S.~V. {Borges}, C.~V. {Rodrigues}, J.~G. {Coelho}, M.~{Malheiro}, and
  M.~{Castro}, ``{A Magnetic White Dwarf Accretion Model for the Anomalous
  X-Ray Pulsar 4U 0142+61},'' {\em \apj}, vol.~895, p.~26, May 2020.

\bibitem{2020MNRAS.498.4426S}
M.~F. {Sousa}, J.~G. {Coelho}, and J.~C.~N. {de Araujo}, ``{Gravitational waves
  from SGRs and AXPs as fast-spinning white dwarfs},'' {\em \mnras}, vol.~498,
  pp.~4426--4432, Nov. 2020.

\bibitem{2020Natur.587...54C}
{CHIME/FRB Collaboration}, B.~C. {Andersen}, K.~M. {Bandura}, M.~{Bhardwaj},
  A.~{Bij}, M.~M. {Boyce}, P.~J. {Boyle}, C.~{Brar}, T.~{Cassanelli},
  P.~{Chawla}, T.~{Chen}, J.~F. {Cliche}, A.~{Cook}, D.~{Cubranic}, A.~P.
  {Curtin}, N.~T. {Denman}, M.~{Dobbs}, F.~Q. {Dong}, M.~{Fandino},
  E.~{Fonseca}, B.~M. {Gaensler}, U.~{Giri}, D.~C. {Good}, M.~{Halpern}, A.~S.
  {Hill}, G.~F. {Hinshaw}, C.~{H{\"o}fer}, A.~{Josephy}, J.~W. {Kania}, V.~M.
  {Kaspi}, T.~L. {Landecker}, C.~{Leung}, D.~Z. {Li}, H.~H. {Lin}, K.~W.
  {Masui}, R.~{McKinven}, J.~{Mena-Parra}, M.~{Merryfield}, B.~W. {Meyers},
  D.~{Michilli}, N.~{Milutinovic}, A.~{Mirhosseini}, M.~{M{\"u}nchmeyer},
  A.~{Naidu}, L.~B. {Newburgh}, C.~{Ng}, C.~{Patel}, U.~L. {Pen},
  T.~{Pinsonneault-Marotte}, Z.~{Pleunis}, B.~M. {Quine}, M.~{Rafiei-Ravandi},
  M.~{Rahman}, S.~M. {Ransom}, A.~{Renard}, P.~{Sanghavi}, P.~{Scholz}, J.~R.
  {Shaw}, K.~{Shin}, S.~R. {Siegel}, S.~{Singh}, R.~J. {Smegal}, K.~M. {Smith},
  I.~H. {Stairs}, C.~M. {Tan}, S.~P. {Tendulkar}, I.~{Tretyakov},
  K.~{Vanderlinde}, H.~{Wang}, D.~{Wulf}, and A.~V. {Zwaniga}, ``{A bright
  millisecond-duration radio burst from a Galactic magnetar},'' {\em \nat},
  vol.~587, pp.~54--58, Nov. 2020.

\bibitem{2020ATel13684....1B}
C.~{Bochenek}, S.~{Kulkarni}, V.~{Ravi}, D.~{McKenna}, G.~{Hallinan}, and
  K.~{Belov}, ``{Independent detection of the radio burst reported in ATel
  \#13681 with STARE2},'' {\em The Astronomer's Telegram}, vol.~13684, p.~1,
  Apr. 2020.

\bibitem{2020ApJ...904L..21Y}
G.~{Younes}, T.~{G{\"u}ver}, C.~{Kouveliotou}, M.~G. {Baring}, C.-P. {Hu},
  Z.~{Wadiasingh}, B.~{Begi{\c{c}}arslan}, T.~{Enoto},
  E.~{G{\"o}{\u{g}}{\"u}{\c{s}}}, L.~{Lin}, A.~K. {Harding}, A.~J. {van der
  Horst}, W.~A. {Majid}, S.~{Guillot}, and C.~{Malacaria}, ``{NICER View of the
  2020 Burst Storm and Persistent Emission of SGR 1935+2154},'' {\em \apjl},
  vol.~904, p.~L21, Dec. 2020.

\bibitem{2021NatAs...5..372R}
A.~{Ridnaia}, D.~{Svinkin}, D.~{Frederiks}, A.~{Bykov}, S.~{Popov},
  R.~{Aptekar}, S.~{Golenetskii}, A.~{Lysenko}, A.~{Tsvetkova}, M.~{Ulanov},
  and T.~L. {Cline}, ``{A peculiar hard X-ray counterpart of a Galactic fast
  radio burst},'' {\em Nature Astronomy}, vol.~5, pp.~372--377, Apr. 2021.

\bibitem{2021NatAs...5..378L}
C.~K. {Li}, L.~{Lin}, S.~L. {Xiong}, M.~Y. {Ge}, X.~B. {Li}, T.~P. {Li}, F.~J.
  {Lu}, S.~N. {Zhang}, Y.~L. {Tuo}, Y.~{Nang}, B.~{Zhang}, S.~{Xiao},
  Y.~{Chen}, L.~M. {Song}, Y.~P. {Xu}, C.~Z. {Liu}, S.~M. {Jia}, X.~L. {Cao},
  J.~L. {Qu}, S.~{Zhang}, Y.~D. {Gu}, J.~Y. {Liao}, X.~F. {Zhao}, Y.~{Tan},
  J.~Y. {Nie}, H.~S. {Zhao}, S.~J. {Zheng}, Y.~G. {Zheng}, Q.~{Luo}, C.~{Cai},
  B.~{Li}, W.~C. {Xue}, Q.~C. {Bu}, Z.~{Chang}, G.~{Chen}, L.~{Chen}, T.~X.
  {Chen}, Y.~B. {Chen}, Y.~P. {Chen}, W.~{Cui}, W.~W. {Cui}, J.~K. {Deng},
  Y.~W. {Dong}, Y.~Y. {Du}, M.~X. {Fu}, G.~H. {Gao}, H.~{Gao}, M.~{Gao}, Y.~D.
  {Gu}, J.~{Guan}, C.~C. {Guo}, D.~W. {Han}, Y.~{Huang}, J.~{Huo}, L.~H.
  {Jiang}, W.~C. {Jiang}, J.~{Jin}, Y.~J. {Jin}, L.~D. {Kong}, G.~{Li}, M.~S.
  {Li}, W.~{Li}, X.~{Li}, X.~F. {Li}, Y.~G. {Li}, Z.~W. {Li}, X.~H. {Liang},
  B.~S. {Liu}, G.~Q. {Liu}, H.~W. {Liu}, X.~J. {Liu}, Y.~N. {Liu}, B.~{Lu},
  X.~F. {Lu}, T.~{Luo}, X.~{Ma}, B.~{Meng}, G.~{Ou}, N.~{Sai}, R.~C. {Shang},
  X.~Y. {Song}, L.~{Sun}, L.~{Tao}, C.~{Wang}, G.~F. {Wang}, J.~{Wang}, W.~S.
  {Wang}, Y.~S. {Wang}, X.~Y. {Wen}, B.~B. {Wu}, B.~Y. {Wu}, M.~{Wu}, G.~C.
  {Xiao}, H.~{Xu}, J.~W. {Yang}, S.~{Yang}, Y.~J. {Yang}, Y.-J. {Yang}, Q.~B.
  {Yi}, Q.~Q. {Yin}, Y.~{You}, A.~M. {Zhang}, C.~M. {Zhang}, F.~{Zhang}, H.~M.
  {Zhang}, J.~{Zhang}, T.~{Zhang}, W.~{Zhang}, W.~C. {Zhang}, W.~Z. {Zhang},
  Y.~{Zhang}, Y.~{Zhang}, Y.~F. {Zhang}, Y.~J. {Zhang}, Z.~{Zhang}, Z.~{Zhang},
  Z.~L. {Zhang}, D.~K. {Zhou}, J.~F. {Zhou}, Y.~{Zhu}, Y.~X. {Zhu}, and R.~L.
  {Zhuang}, ``{HXMT identification of a non-thermal X-ray burst from SGR
  J1935+2154 and with FRB 200428},'' {\em Nature Astronomy}, vol.~5,
  pp.~378--384, Apr. 2021.

\bibitem{2021NatAs...5..401T}
M.~{Tavani}, C.~{Casentini}, A.~{Ursi}, F.~{Verrecchia}, A.~{Addis}, L.~A.
  {Antonelli}, A.~{Argan}, G.~{Barbiellini}, L.~{Baroncelli}, G.~{Bernardi},
  G.~{Bianchi}, A.~{Bulgarelli}, P.~{Caraveo}, M.~{Cardillo}, P.~W. {Cattaneo},
  A.~W. {Chen}, E.~{Costa}, E.~{Del Monte}, G.~{Di Cocco}, G.~{Di Persio},
  I.~{Donnarumma}, Y.~{Evangelista}, M.~{Feroci}, A.~{Ferrari}, V.~{Fioretti},
  F.~{Fuschino}, M.~{Galli}, F.~{Gianotti}, A.~{Giuliani}, C.~{Labanti},
  F.~{Lazzarotto}, P.~{Lipari}, F.~{Longo}, F.~{Lucarelli}, A.~{Magro},
  M.~{Marisaldi}, S.~{Mereghetti}, E.~{Morelli}, A.~{Morselli}, G.~{Naldi},
  L.~{Pacciani}, N.~{Parmiggiani}, F.~{Paoletti}, A.~{Pellizzoni}, M.~{Perri},
  F.~{Perotti}, G.~{Piano}, P.~{Picozza}, M.~{Pilia}, C.~{Pittori},
  S.~{Puccetti}, G.~{Pupillo}, M.~{Rapisarda}, A.~{Rappoldi}, A.~{Rubini},
  G.~{Setti}, P.~{Soffitta}, M.~{Trifoglio}, A.~{Trois}, S.~{Vercellone},
  V.~{Vittorini}, P.~{Giommi}, and F.~{D'Amico}, ``{An X-ray burst from a
  magnetar enlightening the mechanism of fast radio bursts},'' {\em Nature
  Astronomy}, vol.~5, pp.~401--407, Apr. 2021.

\bibitem{2020Natur.587...45Z}
B.~{Zhang}, ``{The physical mechanisms of fast radio bursts},'' {\em \nat},
  vol.~587, pp.~45--53, Nov. 2020.

\bibitem{2020MNRAS.498.1397L}
W.~{Lu}, P.~{Kumar}, and B.~{Zhang}, ``{A unified picture of Galactic and
  cosmological fast radio bursts},'' {\em \mnras}, vol.~498, pp.~1397--1405,
  Oct. 2020.

\bibitem{2010vaoa.conf..129P}
S.~B. {Popov} and K.~A. {Postnov}, ``{Hyperflares of SGRs as an engine for
  millisecond extragalactic radio bursts},'' in {\em Evolution of Cosmic
  Objects through their Physical Activity} (H.~A. {Harutyunian}, A.~M.
  {Mickaelian}, and Y.~{Terzian}, eds.), pp.~129--132, Nov. 2010.

\bibitem{2014MNRAS.442L...9L}
Y.~{Lyubarsky}, ``{A model for fast extragalactic radio bursts.},'' {\em
  \mnras}, vol.~442, pp.~L9--L13, July 2014.

\bibitem{2019MNRAS.485.4091M}
B.~D. {Metzger}, B.~{Margalit}, and L.~{Sironi}, ``{Fast radio bursts as
  synchrotron maser emission from decelerating relativistic blast waves},''
  {\em \mnras}, vol.~485, pp.~4091--4106, May 2019.

\bibitem{2020ApJ...896..142B}
A.~M. {Beloborodov}, ``{Blast Waves from Magnetar Flares and Fast Radio
  Bursts},'' {\em \apj}, vol.~896, p.~142, June 2020.

\bibitem{2017ApJ...843L..26B}
A.~M. {Beloborodov}, ``{A Flaring Magnetar in FRB 121102?},'' {\em \apjl},
  vol.~843, p.~L26, July 2017.

\bibitem{2021JCAP...10..023D}
R.~C. {dos Anjos}, J.~G. {Coelho}, J.~P. {Pereira}, and F.~{Catalani},
  ``{High-energy gamma-ray emission from SNR G57.2+0.8 hosting SGR
  J1935+2154},'' {\em \jcap}, vol.~2021, p.~023, Oct. 2021.

\bibitem{Hunter_1997}
S.~D. Hunter, D.~L. Bertsch, J.~R. Catelli, T.~M. Dame, S.~W. Digel, B.~L.
  Dingus, J.~A. Esposito, C.~E. Fichtel, R.~C. Hartman, G.~Kanbach, D.~A.
  Kniffen, Y.~C. Lin, H.~A. Mayer-Hasselwander, P.~F. Michelson, C.~von
  Montigny, R.~Mukherjee, P.~L. Nolan, E.~Schneid, P.~Sreekumar, P.~Thaddeus,
  and D.~J. Thompson, ``{EGRET} observations of the diffuse gamma-ray emission
  from the galactic plane,'' {\em The Astrophysical Journal}, vol.~481,
  pp.~205--240, may 1997.

\bibitem{Fermi2012}
M.~{Ackermann} and et~al., ``{Fermi-LAT Observations of the Diffuse
  {\ensuremath{\gamma}}-Ray Emission: Implications for Cosmic Rays and the
  Interstellar Medium},'' {\em \apj}, vol.~750, p.~3, May 2012.

\bibitem{HESS2006}
F.~{Aharonian} and et~al., ``{Discovery of very-high-energy
  {\ensuremath{\gamma}}-rays from the Galactic Centre ridge},'' {\em \nat},
  vol.~439, pp.~695--698, Feb. 2006.

\bibitem{Archer_2016}
A.~{Archer} and et~al., ``{TeV Gamma-Ray Observations of the Galactic Center
  Ridge by VERITAS},'' {\em \apj}, vol.~821, p.~129, Apr. 2016.

\bibitem{Magic2020}
{MAGIC Collaboration}, ``{MAGIC observations of the diffuse
  {\ensuremath{\gamma}}-ray emission in the vicinity of the Galactic center},''
  {\em \aap}, vol.~642, p.~A190, Oct. 2020.

\bibitem{HESS2016}
{HESS Collaboration}, ``{Acceleration of petaelectronvolt protons in the
  Galactic Centre},'' {\em \nat}, vol.~531, pp.~476--479, Mar. 2016.

\bibitem{2020ApJ...902L..43L}
L.~{Lin}, E.~{G{\"o}{\u{g}}{\"u}{\c{s}}}, O.~J. {Roberts}, M.~G. {Baring},
  C.~{Kouveliotou}, Y.~{Kaneko}, A.~J. {van der Horst}, and G.~{Younes},
  ``{Fermi/GBM View of the 2019 and 2020 Burst Active Episodes of SGR
  J1935+2154},'' {\em \apjl}, vol.~902, p.~L43, Oct. 2020.

\bibitem{2021ApJ...919..106A}
{H.~E.~S.~S. Collaboration}, ``{Searching for TeV Gamma-Ray Emission from SGR
  1935+2154 during Its 2020 X-Ray and Radio Bursting Phase},'' {\em \apj},
  vol.~919, p.~106, Oct. 2021.

\bibitem{Strong_1998}
A.~W. Strong and I.~V. Moskalenko, ``Propagation of cosmic-ray nucleons in the
  galaxy,'' {\em The Astrophysical Journal}, vol.~509, pp.~212--228, dec 1998.

\bibitem{Porter_2017}
T.~A. Porter, G.~J{\'{o}}hannesson, and I.~V. Moskalenko, ``High-energy gamma
  rays from the milky way: Three-dimensional spatial models for the cosmic-ray
  and radiation field densities in the interstellar medium,'' {\em The
  Astrophysical Journal}, vol.~846, p.~67, aug 2017.

\bibitem{J_hannesson_2018}
G.~J{\'{o}}hannesson, T.~A. Porter, and I.~V. Moskalenko, ``The
  three-dimensional spatial distribution of interstellar gas in the milky way:
  Implications for cosmic rays and high-energy gamma-ray emissions,'' vol.~856,
  p.~45, mar 2018.

\bibitem{2021arXiv211212745P}
T.~A. {Porter}, G.~{Johannesson}, and I.~V. {Moskalenko}, ``{The GALPROP
  Cosmic-ray Propagation and Non-thermal Emissions Framework: Release v57},''
  {\em arXiv e-prints}, p.~arXiv:2112.12745, Dec. 2021.

\bibitem{2019ApJ...879...91J}
G.~{J{\'o}hannesson}, T.~A. {Porter}, and I.~V. {Moskalenko}, ``{Cosmic-Ray
  Propagation in Light of the Recent Observation of Geminga},'' {\em \apj},
  vol.~879, p.~91, July 2019.

\bibitem{2020ApJ...905...99Z}
P.~{Zhou}, X.~{Zhou}, Y.~{Chen}, J.-S. {Wang}, J.~{Vink}, and Y.~{Wang},
  ``{Revisiting the Distance, Environment, and Supernova Properties of SNR
  G57.2+0.8 that Hosts SGR 1935+2154},'' {\em \apj}, vol.~905, p.~99, Dec.
  2020.

\bibitem{HESS2018}
{H.~E.~S.~S. Collaboration}, ``{Characterising the VHE diffuse emission in the
  central 200 parsecs of our Galaxy with H.E.S.S.},'' {\em \aap}, vol.~612,
  p.~A9, Apr. 2018.

\bibitem{Tibet2021}
M.~{Amenomori} and et~al., ``{First Detection of sub-PeV Diffuse Gamma Rays
  from the Galactic Disk: Evidence for Ubiquitous Galactic Cosmic Rays beyond
  PeV Energies},'' {\em Physical Review Letters}, vol.~126, p.~141101, Apr.
  2021.

\bibitem{2009PhRvD..80f3005M}
D.~{Malyshev}, I.~{Cholis}, and J.~{Gelfand}, ``{Pulsars versus dark matter
  interpretation of ATIC/PAMELA},'' {\em \prd}, vol.~80, p.~063005, Sept. 2009.

\bibitem{2011A&A...536A..83S}
X.~H. {Sun}, P.~{Reich}, W.~{Reich}, L.~{Xiao}, X.~Y. {Gao}, and J.~L. {Han},
  ``{A Sino-German {\ensuremath{\lambda}}6 cm polarization survey of the
  Galactic plane. VII. Small supernova remnants},'' {\em \aap}, vol.~536,
  p.~A83, Dec. 2011.

\bibitem{2013ApJS..204....4P}
M.~Z. {Pavlovi{\'c}}, D.~{Uro{\v{s}}evi{\'c}}, B.~{Vukoti{\'c}}, B.~{Arbutina},
  and {\"U}.~D. {G{\"o}ker}, ``{The Radio Surface-brightness-to-Diameter
  Relation for Galactic Supernova Remnants: Sample Selection and Robust
  Analysis with Various Fitting Offsets},'' {\em Astrophysical Journal
  Supplement Series}, vol.~204, p.~4, Jan. 2013.

\end{thebibliography}




\end{document}